\documentclass[12pt]{iopart}

\usepackage{iopams}  
\usepackage{graphicx}
\usepackage{siunitx}
\newcommand{\abs}[1]{\left| #1 \right|}

\usepackage{xcolor,soul} %used for highlighting
\begin{document}

\title{{Multi-twist polarization ribbon topologies in highly-confined optical fields}}

\author{Thomas Bauer$^{1,2,3}$, Peter Banzer$^{1,2,4}$, Fr\'ed\'eric Bouchard$^{4}$, Sergej Orlov$^{5}$, Lorenzo Marrucci$^{6}$, Enrico Santamato$^{6}$, Robert W. Boyd$^{1,4,7}$, Ebrahim Karimi$^{1,4}$, Gerd Leuchs$^{1,2,4}$}

\address{$^{1}$ Max Planck Institute for the Science of Light, Staudtstr. 2, D-91058 Erlangen, Germany}
\address{$^{2}$ Institute of Optics, Information and Photonics, Friedrich-Alexander-University Erlangen-Nuremberg, Staudtstr. 7/B2, D-91058 Erlangen, Germany}
\address{$^{3}$ Kavli Institute of Nanoscience Delft, Delft University of Technology, Lorentzweg 1, Delft 2628 CJ, The Netherlands}
\address{$^{4}$ Department of Physics, University of Ottawa, 25 Templeton Street, Ottawa, Ontario, K1N 6N5 Canada}
\address{$^{5}$ State Research Institute Center for Physical Sciences and Technology, Industrial Laboratory for Photonic Technologies, Sauletekio ave 3 LT-10222, Vilnius, Lithuania}
\address{$^{6}$ Dipartimento di Fisica ``Ettore Pancini'', Universit\'a di Federico II, Compl. Univ. Monte S. Angelo, via Cintia, 80126 Napoli, Italy}
\address{$^{7}$ Institute of Optics, University of Rochester, Rochester, New York, 14627, USA}
\ead{peter.banzer@mpl.mpg.de, ekarimi@uottawa.ca}
\vspace{10pt}
%\begin{indented}
%\item[]August 2017
%\end{indented}

\begin{abstract}
Electromagnetic plane waves, solutions to Maxwell's equations, are said to be ``transverse'' in vacuum. Namely, the waves' oscillatory electric and magnetic fields are confined within a plane transverse to the waves' propagation direction. Under tight-focusing conditions {however}, the {field can exhibit} longitudinal electric or magnetic components, \emph{transverse} spin angular momentum, or non-trivial topologies such as M\"obius strips. Here, we show that when a {suitably} spatially structured beam is tightly focused, a 3-dimensional polarization topology in the form of a ribbon with two full twists appears in the focal volume. We study experimentally the stability and dynamics of the observed polarization ribbon by exploring its topological structure for various radii upon focusing and for different propagation planes. 
\end{abstract}

%
% Uncomment for keywords
%\vspace{2pc}
%\noindent{\it Keywords}: XXXXXX, YYYYYYYY, ZZZZZZZZZ
%
% Uncomment for Submitted to journal title message
%\submitto{\JPA}
%
% Uncomment if a separate title page is required
%\maketitle
% 
% For two-column output uncomment the next line and choose [10pt] rather than [12pt] in the \documentclass declaration
%\ioptwocol
%

%%%%%%%%%%%%%%%%%%%%%%%%%%%%%%
\section{Introduction}
Since the inception of electromagnetic theory, the polarization of light, i.e. the oscillation direction of the electric field vector, has been a central concept to our understanding of optics, giving rise to countless applications~\cite{born:13}. For plane waves and in paraxial beams, the polarization has been recognized as a transverse quantity and, hence, it can be represented by a set of two orthogonal basis vectors. For instance, in the linear and circular bases, the polarization of an optical beam can be represented by superpositions of linearly horizontally and linearly vertically, or circularly-left and circularly-right polarized beams, respectively. The ratio and the relative phase between the two polarization components define the oscillations of the electric field vector's tip upon propagation or in time, and its trajectory in the plane transverse to its propagation direction, typically given by an ellipse~\cite{born:13}. {This description of the light field by a so-called polarization ellipse at each point in space is even valid in highly confined fields exhibiting out-of-plane field components, as long as the field itself is monochromatic.} In two different cases, this ellipse becomes \emph{singular}~\cite{nye:83,berry:04,nye:87}: (i) the ellipse's major and minor axes are undefined, resulting in circular polarization (C-point); (ii) the minor axis of the ellipse is zero and its surface normal is undefined, and thus the polarization is linear (L-line). 
%Optical beams possessing uniform polarization distributions in the plane transverse to their propagation direction have been extensively studied and widely exploited.
{These so-called polarization singularities in general arise in light fields with} spatially inhomogeneous polarization distributions, which we refer to as space-varying polarized light beams. They have recently received great attention {owing} to their peculiar optical features and applications~\cite{zhan:09,rubinsztein:16,aiello:15}. Vector vortex beams~\cite{zhan:09} - space-varying linearly-polarized beams - and Poincar\'e beams~\cite{beckley:10} - optical beams containing all types of polarization - are among this class of spatially inhomogeneously polarized beams. These beams are interesting both at the fundamental and applied levels. For instance, they can be used to enhance measurement sensitivity~\cite{d:13,berg:15}, to transport high-power in nonlinear media~\cite{bouchard:16}, or to generate exotic optical beams with peculiar topological structures~\cite{otte:16,larocque:18}. {In particular, spatially structured light fields with field components along the propagation direction were predicted by Freund to show so-called optical polarization M\"obius strips and twisted ribbons}~\cite{freund:05,freund:10a}{, with the former recently confirmed experimentally in tightly focused fields}~\cite{bauer:15} {as well as the originally proposed scheme of crossing beams}~\cite{galvez:17}. {Here, we will experimentally demonstrate the generation and stability of the latter in highly confined fields, specifically looking at the dynamics of the twisted ribbon when propagating through the focal volume of a tailored space-varying polarized light beam.}

%In particular, radially and azimuthally polarized beams - so-called cylindrical vector beams - which carry no net optical angular momentum, generate \emph{needle{-like}} electric and magnetic {fields}, respectively, oriented along the propagation direction~\cite{quabis:00,dorn:03,youngworth:00}, {and} may be used to excite electric dipole atomic transitions with high efficiency~\cite{sondermann:07,lindlein:07}. {In addition, such light beams are also a versatile tool} for the selective excitation of individual nanostructures~\cite{kindler:07,wozniak:15}.

%%%%%%%%%%%%%%%%%%%%%%%%%%%%%%
\section{Space-varying polarized beams under tight focusing}
\begin{figure}[t]
	\centering
	\includegraphics[width=0.6\columnwidth]{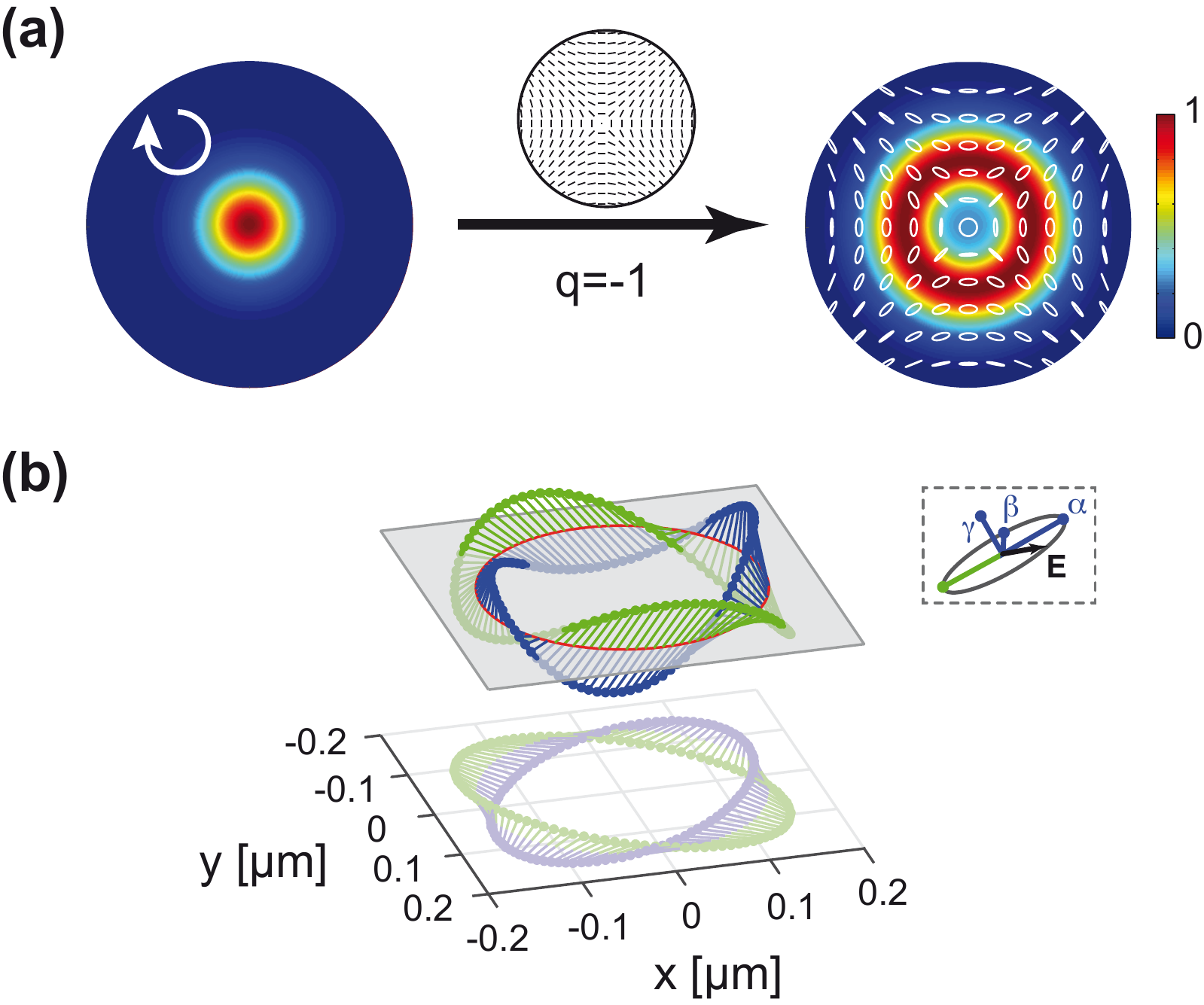}
	\caption{\textbf{Creation of a full Poincar\'e beam and subsequent optical polarization ribbon with twist index $-2$ under tight focusing.} \textbf{(a)} A left-handed circularly polarized Gaussian beam is converted into a full Poincar\'e beam by means of a space-variant birefringent plate ($q$-plate) with topological charge $q=-1$ (see sketch for the orientation of its fast axes) and optical retardation of $\delta=2.92\,\text{rad}$. The resulting polarization ellipse at each point in the transverse plane of the beam is superimposed in white. \textbf{(b)} Twisted ribbon with 2 twists and its projection onto the transverse plane, created by tight focusing of the beam shown in \textbf{(a)} and tracing the major axis $\alpha$ of the polarization ellipse (see inset for a sketch with the definition of its parameters) on a circle with $\rho=150~\text{nm}$ around the optical axis.}%Major axis of the polarization ellipse traced on a circle with $\rho=150~\text{nm}$ around the optical axis for three different transverse planes; before the focus $z=-\lambda$, at the focus $z=0$, and after the focus $z=+\lambda$, and \textbf{(b)} their projections onto the transverse plane. The results associated to the planes of $z=\pm\lambda$ are numerically simulated, and the inset shows a sketch of the polarization ellipse with the definition of its axes.}
	\label{fig:ribbon_propagation}
\end{figure}

There are several different methods to generate optical beams possessing inhomogeneous polarization distributions {in their transverse plane}. For instance, phase-only spatial light modulators (SLMs)~\cite{maurer:07,galvez:12,han:13}, non-unitary polarization transformation~\cite{sit:17}, and spatially structured birefringent plates~\cite{beckley:10,cardano:12} %,larocque:16}
have so far been used to generate Poincar\'e or vector vortex beams. In this article, we use the latter technique to generate full Poincar\'e beams~\cite{beckley:10} by means of a spatially structured liquid crystal device, referred to as $q$-plate~\cite{marrucci:06}. The $q$-plate couples {spin to orbital} angular momentum, and thus allows for generating certain classes of space-varying optical beams {when} it is fed with an elliptically polarized input beam. {Choosing a left-handed circularly polarized Gaussian beam, $\hat{\mathbf e}_\text{L}$, as input, the $q$-plate coherently transforms this field distribution into}
%However, when its optical retardation is chosen to be $\delta$, it coherently transfers a left-hand circularly polarized Gaussian beam, $\hat{\mathbf e}_\text{L}$, into
%
\begin{eqnarray}\label{eq:q-plate}
	\mathbf{E}(\mathbf r) \approx \left(\cos{\left(\frac{\delta}{2}\right)}\,E_\text{L}(\rho;z)\,\hat{\mathbf e}_\text{L}+i\sin{\left(\frac{\delta}{2}\right)}\,E_\text{R}(\rho;z)\,e^{2iq\phi}\, \hat{\mathbf e}_\text{R}\right),\!\!\!\!
\end{eqnarray}
where $\delta$ is the chosen optical retardation of the $q$-plate, $\rho,\phi,z$ are cylindrical coordinates, $q$ is the $q$-plate's topological charge, $\hat{\mathbf e}_\text{L}$ and $\hat{\mathbf e}_\text{R}$ are the left- and right-handed polarization unit vectors, and $E_\text{L}(\rho;z)$ and $E_\text{R}(\rho;z)$ are the beam profiles of the left- and right-handed circularly polarized beams, respectively~\cite{cardano:13}. The orientation of liquid crystal molecules in the transverse plane of a given $q$-plate with topological charge of {$q=-1$} is shown in Fig.~\ref{fig:ribbon_propagation}-\textbf{(a)}. Adjusting the optical retardation $\delta$ of the plate changes the superposition {ratio}, and thus the polarization topology. For instance, when (i) $\delta=0, 2\pi$, the plate won't change the initial state of the beam and the beam polarization remains left-handed circular; (ii) $\delta=\pi$, the plate works as a structured half-wave plate and converts the Gaussian beam into a right-handed circularly polarized doughnut beam possessing orbital angular momentum of $2q$; (iii) $\delta=\pi/2$ the plate generates a coherent superposition of (i) and (ii). In the latter case (iii), the beam exhibits a point of circular polarization on the optical axis (C-point) and an azimuthal polarization structure with polarization topological charge of $\eta=q$. An example of such polarization topology is shown in Fig.~\ref{fig:ribbon_propagation}-\textbf{(a)}. Note that the polarization topology shown in Fig.~\ref{fig:ribbon_propagation}-\textbf{(a)} corresponds to that resulting from a $q=-1$-plate. However, for practical reasons we use a $q=1$-plate followed by a half-wave plate in the experiment to mimic the $q=-1$-plate topology (see Fig.~\ref{fig:setup}).\newline

Upon tight focusing, strong (longitudinal) $z$-components of the incoming transversely polarized beam described by Eq.~(\ref{eq:q-plate}) emerge. Using vectorial diffraction theory~\cite{richards:59}, one can show that the total electric field at the focus will have the following form~\cite{kostya:11}
\begin{eqnarray}\label{eq:focus}
	\mathbf{E}_\text{focus}(\mathbf r)&=&\left(\widetilde{E}_\text{L}(\rho;z)\,\hat{\mathbf e}_\text{L}+\widetilde{E}_\text{R}(\rho;z)\,e^{2iq\phi}\, \hat{\mathbf e}_\text{R}\right)\cr\cr
	&+&\left(\widetilde{E}^{z}_\text{L}(\rho;z) e^{i\phi}+\widetilde{E}^{z}_\text{R}(\rho;z)e^{i(2q-1)\phi}\right)\hat{\mathbf{e}}_\text{z},
\end{eqnarray}
where $\widetilde{E}_\text{L}$ and $\widetilde{E}_\text{R}$ are the components of the transverse electric field (including the weight factors) at the focus, and $\widetilde{E}^{z}_\text{L}$ and $\widetilde{E}^{z}_\text{R}$ are the associated $z$-components. The exact form for these amplitudes, i.e. $\widetilde{E}_\text{L}$, $\widetilde{E}_\text{R}$, $\widetilde{E}^{z}_\text{L}$ and $\widetilde{E}^{z}_\text{R}$, can be found in~\cite{kostya:11}. The last term in Eq.~(\ref{eq:focus}) corresponds to spin-orbit coupling {amplified by tight focusing with a} high-numerical aperture lens, which renders the polarization vector in 3-dimensional space, {with} its amplitude {depending} on the numerical aperture (NA) of the focusing lens. However, the polarization ellipse traced by the tip of the electric field vector in time remains in a 2-dimensional plane for any given point {in space}. Nevertheless, {the spatial distribution of} the polarization ellipse forms a specific topology in 3-dimensional space, which is dictated by the topological charge of the $q$-plate. Without loss of generality, we consider a radial position $\rho=\rho_0$ {where the amplitude of the contributions of the initial right- and left-circular polarization components to the longitudinal focal field component are equal}, i.e. $\widetilde{E}^{z}_\text{L}(\rho_0)=\widetilde{E}^{z}_\text{R}(\rho_0)$. The $z$-component of the electric field, apart from a phase $\exp{(iq\phi)}$, will now be proportional to $\cos \left[ (q-1) \phi \right]$. The electric field intensity of the $z$-component will then be proportional to $\left| \cos \left[ (q-1) \phi \right] \right|^2$, which results in a $2 \left| q-1 \right|$-fold symmetry. {This} z-component of the electric field turns the 2-dimensional into a 3-dimensional polarization topology. The structure of the polarization topology can be studied by evaluating the spatial dependence {of} either the major or minor axes of the polarization ellipse. {Tracing the focal field on a closed loop} around the on-axis C-point, the major axis of the polarization ellipse oscillates $|q-1|$ times {through the transverse plane}. In the 3-dimensional perspective, the major (or the minor) axis of the polarization ellipse, depending on the value of $q$, forms a M\"obius or a ribbon topology with $|q-1|$ twists. {While} the existence of polarization M\"obius strips with 3/2-twists and 5/2-twists, for the case of $q=-1/2$ and $-3/2$, respectively, has been experimentally demonstrated recently~\cite{bauer:15}, {we here will look at the case of ribbons with integer twists for integer values of $q$.}

%%%%%%%%%%%%%%%%%%%%%%%%%%%%%%
\section{Experimental Realization}
To determine the polarization ellipse, and consequently {its} major and minor axes in three dimensions, it is necessary to measure amplitude and phase of the {full} complex vectorial light field in the plane of observation. This can be achieved by employing a probe as a local field sensor and utilizing the angularly resolved far-field scattering by this probe while point-wise scanning the latter relative to the field distribution. This technique, named Mie-scattering nano-interferometry (see~\cite{bauer:14} for more details on the technique), was shown to achieve deep sub-wavelength spatial resolution in the experimental study of vectorial focal fields. The Mie-scattering nano-interferometry has previously allowed for the experimental verification of optical polarization M\"obius strips in both, tailored light fields~\cite{bauer:15} and in the focal region of a tightly focused linearly polarized beam around points of transversely spinning fields~\cite{bauer:16}. The general experimental concept will be discussed in the following, while details are discussed in~\cite{bauer:14}.

%%%%%
\begin{figure}[t]
	\centering
	\includegraphics[width=0.75\columnwidth]{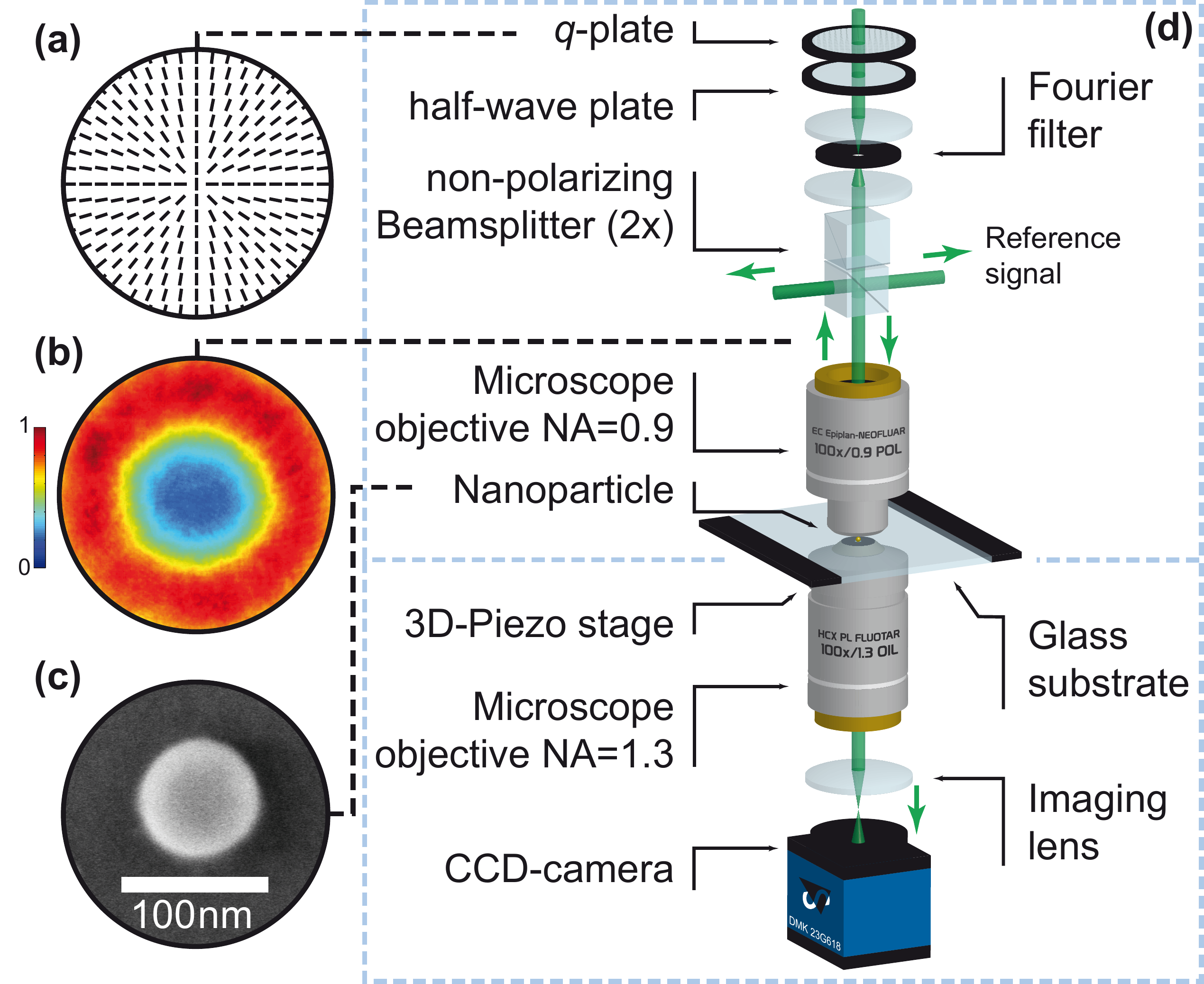}
	\caption{\textbf{Sketch of the experimental setup.} \textbf{(a)} Distribution of the local fast axis of the liquid crystal molecules in the transverse plane of the employed $q=1$-plate. \textbf{(b)} {Experimentally generated intensity distribution of the resulting full-Poincar\'e beam that is subsequently focused. Note} that a half-wave plate is used to transform the output polarisation topology into that of a $q=-1$-plate. \textbf{(c)} SEM-image of the utilized gold nano-sphere with a diameter of $\SI{80}{nm}$. \textbf{(d)} Sketch of the experimental setup utilized to reconstruct the full vectorial focal field distribution of the tightly focused full Poincar\'e beam.}
	\label{fig:setup}
\end{figure}
%%%%%

The custom-built experimental setup is shown in Fig.~\ref{fig:setup}~\cite{bauer:14}. As a nano-probe, a single gold sphere with a transverse diameter of $d=80~\text{nm}$ and a height of $h=86~\text{nm}$ (see scanning electron micrograph in Fig.~\ref{fig:setup}-\textbf{(c)}) on a glass substrate is utilized. For the measurement, it is moved through the investigated field distribution via a 3-dimensional piezo stage. The angularly resolved detection of the interference between the light scattered by the probe into the lower half-space forward direction and the directly transmitted light in {this} region is realized by collecting the light via an oil-immersion microscope objective ($\text{NA}=1.3$) and imaging its back focal plane onto a CCD camera (see Fig. \ref{fig:setup}-\textbf{(d)}). This interferometric information for each position of the probe relative to the investigated field is equivalent to an observation of the scattering process from various directions. Thus, it allows for a retrieval of the relative phase information of the distributions under study from the far-field~\cite{bauer:14}. The highly confined field distribution containing {a twisted ribbon formed by tracing the major axis of the polarization ellipse along a closed loop around the optical axis} is created in the shown setup by sending an initially right-handed circularly polarized Gaussian beam onto a $q=1$-plate \cite{marrucci:06} (see sketch of its structure in Fig.~\ref{fig:setup}-\textbf{(a)}) and a subsequent half-wave plate. 

%%%%%%%%%
\begin{figure*}[t]
	\centering
	\includegraphics[width=\columnwidth]{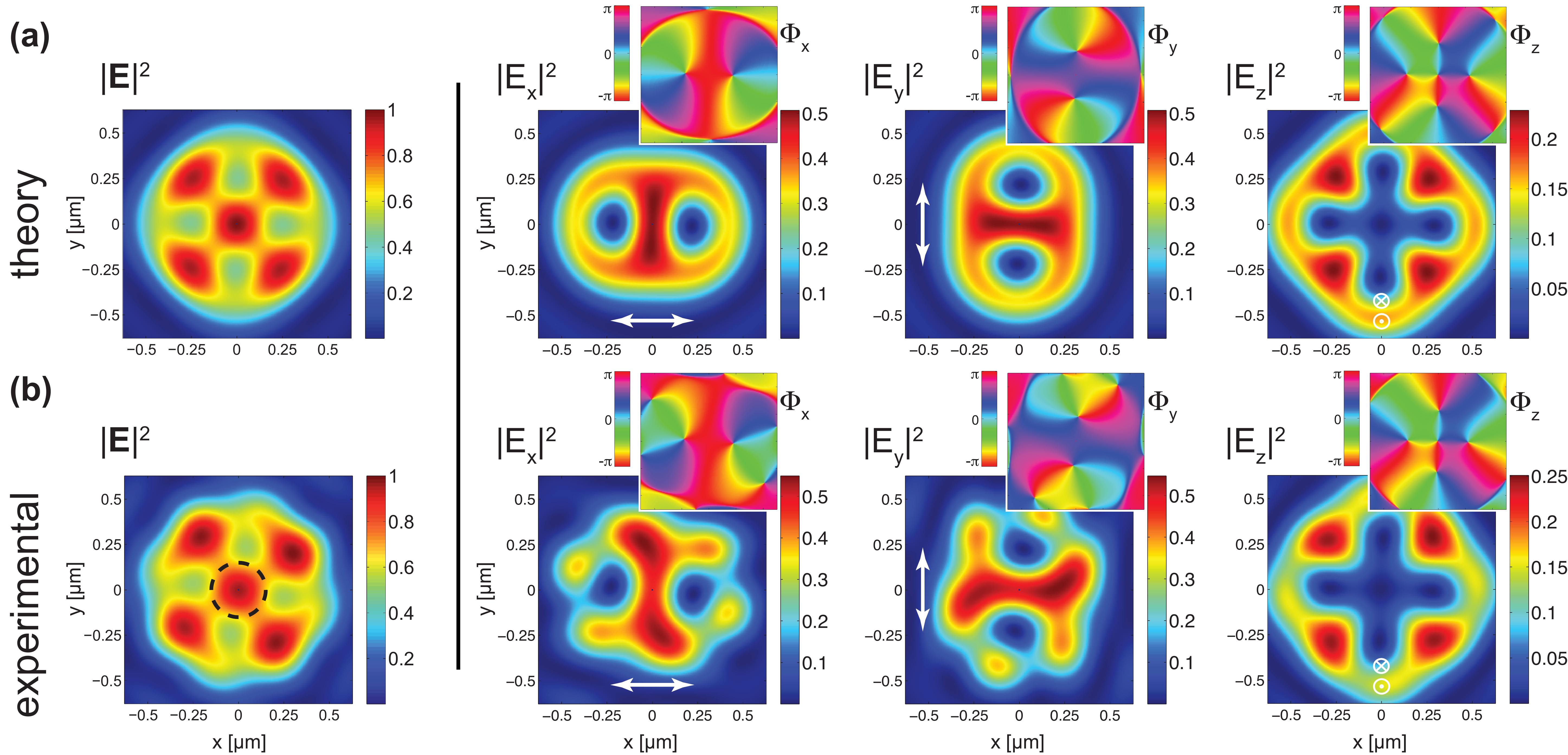}
	\caption{\textbf{Electric energy density distributions of a tightly focused structured beam.} \textbf{(a)} Numerically calculated focal field distribution for a tightly focused composite beam generated from a coherent superposition of $\text{LG}_{0,0}(\rho,\phi,z)\, \hat{\mathbf e}_\text{R}$ and $\text{LG}_{0,2}(\rho,\phi,z)\, \hat{\mathbf e}_\text{L}$. The total electric energy density distribution is plotted on the left, while the individual Cartesian components as well as their relative phase distribution are depicted on the right. \textbf{(b)} Experimentally reconstructed focal field distribution of the same input field used in \textbf{(a)}, showing the very good overlap between experiment and calculation. All distributions are normalized to the maximum value of the corresponding total electric energy density distribution.}
	\label{fig:exp_ribbon_field}
\end{figure*}
%%%%%%%%%

As discussed in the previous section, this combination results effectively in the operation of {a} $q=-1$-plate. By adjusting the voltage applied to the liquid-crystal-based $q$-plate~\cite{cardano:13,larocque:16}, a coaxial superposition of the initial right-handed circular polarized Gaussian beam ($\text{HyGG}_{0,0}(\rho,\phi,z)\, \hat{\mathbf e}_\text{R}$) and a left-handed circularly polarized hypergeometric Gauss beam ($\text{HyGG}_{-2,2}(\rho,\phi,z)\,\hat{\mathbf e}_\text{L}$) is generated~\cite{karimi:09,karimi:07} with the Hypergeometric Gauss modes $\text{HyGG}_{p,\ell}(\rho,\phi,z)$ having the radial and azimuthal indices of $p$ and $\ell$, respectively. The resulting full Poincar\'e beam~\cite{beckley:10,cardano:13} is filtered spatially with a pinhole to obtain the lowest radial order of both constituting beams in the Laguerre Gauss basis, i.e. $\text{LG}_{0,0}(\rho,\phi,z)\,\hat{\mathbf e}_\text{R}$ and $\text{LG}_{0,2}(\rho,\phi,z) \,\hat{\mathbf e}_\text{L}$ {(see} Fig.~\ref{fig:setup}-\textbf{(b)} {for the experimentally achieved intensity distribution)}. The spatially filtered beam is then transmitted through two orthogonally oriented non-polarizing beamsplitters to redirect part of the incoming beam and the light {reflected from the sample} onto corresponding photodetectors. By using two orthogonally aligned beamsplitters, the remaining weak polarizing effect of non-polarizing beamsplitters can be compensated for. Finally, the generated beam is focused by a microscope objective with an NA of $0.9$, resulting in the complex focal field distribution under study, {shown in Fig.~}\ref{fig:exp_ribbon_field}-\textbf{(a)}.

Scanning the described nano-probe through this focal field and applying the reconstruction algorithm~\cite{bauer:14} to the collected far-field intensity information results in the experimentally reconstructed focal field distributions shown in Fig.~\ref{fig:exp_ribbon_field}-\textbf{(b)}. Here, the excitation wavelength was chosen to be $\lambda=530~\text{nm}$, with an experimentally determined relative permittivity of the utilized nano-probe of $\epsilon= -3.1 + 2.5\,i$. The total electric energy density (depicted on the left side of Fig.~\ref{fig:exp_ribbon_field}-\textbf{(b)}) strongly resembles the numerically simulated field distributions calculated via vectorial diffraction theory (Fig.~\ref{fig:exp_ribbon_field}-\textbf{(a)})~\cite{richards:59,novotny:06}. The energy density distributions of the individual electric field components (right side of Fig.~\ref{fig:exp_ribbon_field}-\textbf{(b)}) show minor deviations specifically in the transverse field components. The resulting phase distributions are also shown {as} insets. The phase distribution of the transverse components of the electric field at the focus exhibit two singularities of topological charge $\pm1$, both displaced (vertically or horizontally) away from the optical axis. Due to spin-orbit coupling, the $z$-component of the electric field under tight focusing gains extra phase singularity points, in this case five singular points that are shown in Fig.~\ref{fig:exp_ribbon_field}. The $z$-component of the electric field now reaches amplitudes comparable to that of the transverse components (see the scale bar in Fig.~\ref{fig:exp_ribbon_field}-\textbf{(b)}), and thus breaks the cylindrical symmetry into a four-fold symmetric pattern. 
%%%%%%%%%
\begin{figure}[t]
	\centering
	\includegraphics[width=0.65\columnwidth]{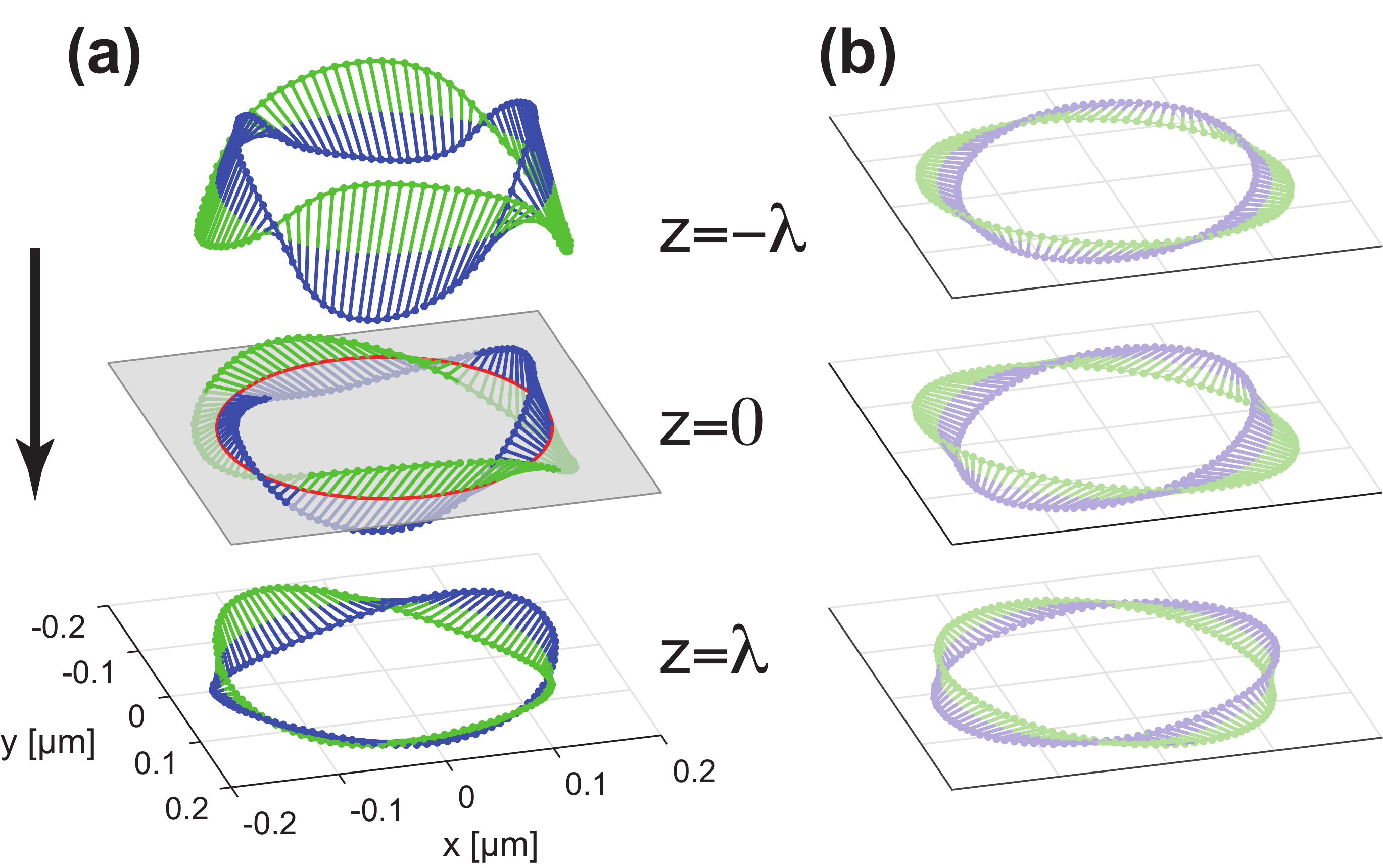}
	\caption{\textbf{Experimentally reconstructed major axes of the polarization ellipses {in the focal volume}.} Experimentally reconstructed distribution of the major axis of the polarization ellipse traced around the optical axis in the focal plane of the tightly focused structured beam as well as one wavelength before and after the focus. \textbf{(a)} The resulting polarization ribbon with twist number of $-2$ and \textbf{(b)} its projection onto the transverse plane are depicted for a trace radius of $\rho=150~\text{nm}$, shown as a red solid line in the semi-transparent focal plane.}
	\label{fig:exp_ribbon_ellipse}
\end{figure}
%%%%%%%%% colour-coded analog to Fig.~\ref{fig:ribbon_propagation}

The major and minor semi-axes of the polarization ellipse $\boldsymbol{\alpha}(\mathbf r),\boldsymbol{\beta}(\mathbf r)$ as well as the normal to the polarization ellipse $\boldsymbol{\gamma}(\mathbf r)$ for the  electric field $\mathbf{E}$ at any point of $\mathbf r$ can be calculated by~\cite{berry:04,dennis:02} 
\begin{eqnarray}\label{eq:mmn_ellipse}
  \begin{array}{l}
  \boldsymbol{\alpha}(\mathbf r)=\frac{1}{\abs{\sqrt{\mathbf{E}\cdot\mathbf{E}}}}\,\Re\left(\mathbf{E}\sqrt{\mathbf{E}^{*}\cdot\mathbf{E}^{*}}\right),\cr 
  \boldsymbol{\beta}(\mathbf r)=\frac{1}{\abs{\sqrt{\mathbf{E}\cdot\mathbf{E}}}}\,\Im\left(\mathbf{E}\sqrt{\mathbf{E}^{*}\cdot\mathbf{E}^{*}}\right)\cr 
  \boldsymbol{\gamma}(\mathbf r)=\Im\left(\mathbf{E}^{*}\times\mathbf{E}\right), 
  \end{array},
\end{eqnarray}
where $\Re(u)$, $\Im(u)$ and $u^*$ represent the real and imaginary parts of $u$, and its complex conjugate, respectively. With these equations, we calculated the local polarization ellipse in the focal plane, $z=0$, from the experimental field data at each point on a circular trace with radius $\rho=150~\text{nm}$ around the optical axis (shown in Fig.~\ref{fig:exp_ribbon_field}-\textbf{(b)} as a black dashed circle). The major axes of the polarization ellipses for these points in 3-dimensional space are shown in {the central row of} Fig.~\ref{fig:exp_ribbon_ellipse}. In order to see the 3-dimensional topological structure, the semi-axes are coloured in blue and green, revealing a twisted ribbon with twist index $-2$. Note that the number of twists is given by $|q-1|$, which for the above case is $|-1-1|=2$ (the minus sign indicates that the direction of the twists is clockwise). The projection of the major axis onto the transverse plane is shown {next to} the ribbon in Fig.~\ref{fig:exp_ribbon_ellipse}-\textbf{(b)}. Following the major axes of the ellipses around the C-point also shows the 2-dimensional polarization topology with the polarization topological index of $-1$ in the transverse plane. We observe the same 3-dimensional (ribbon with two twists) and 2-dimensional (polarization topological index of $-1$) topologies for different radii around the C-point (not shown here). However, for radii, $\rho\gg500$~nm, the field amplitudes drop quickly due to the strong spatial confinement of the focal field, resulting in a low signal-to-noise ratio.

In a next step, we also studied the behavior of the observed ribbon topology in other planes parallel to the focal plane within the focal volume. In order to observe the evolution of the 3-dimensional polarization topology, we retrieve the electric field components and their relative phases one wavelength before and after the focus, i.e. $z=\pm\lambda=\pm530~\text{nm}$ from the reconstructed focal data. Again, the major axes of the polarization ellipses are retrieved for a given radius, i.e. $\rho=150~\text{nm}$, and plotted in 3-dimensional space. Figure~\ref{fig:exp_ribbon_ellipse} shows the evolution of the major axes of the polarization ellipses upon free-space propagation when it passes through the focal plane. The 3-dimensional polarization topology, i.e. ribbon with $-2$ twists, as well as the 2-dimensional polarization topology, shown in Fig.~\ref{fig:exp_ribbon_ellipse}-\textbf{(b)} as a projection onto the corresponding planes, are conserved upon propagation through the focus. However, two main effects can be observed. First, the magnitude of the $z$-component of electric field is weaker outside the focal plane. Second, the topological structure rotates while traversing the focal plane. The latter effect is more visible in the projection shown in Fig. 4-(b) as 2-dimensional topology, and was previously observed for the 2-dimensional case~\cite{cardano:13}. Such a rotation in the 3-dimensional and 2-dimensional polarization topologies is caused by the difference in Gouy phases for $\text{LG}_{0,0}(\rho,\phi,z)$ and $\text{LG}_{0,-2}(\rho,\phi,z)$ beams. This propagation distance-dependent phase equals $-(2p+|\ell|+1)\arctan{(z/z_R)}$ for $\text{LG}_{p,\ell}(\rho,\phi,z)$, where $z_R$ is the Rayleigh range. Thus, one expects a relative accumulated phase when propagating from $-z$ to the focal plane (with $|z| >> z_R$), and, hence, a rotation of $\pi/2$ of the polarization topology.

%%%%%%%%%%%%%%%%%%%%%%%%%%%%%%
\section{Conclusion}
In summary, we {studied the topological structure of} an optical beam possessing a transverse polarization topological charge of $-1$ {in the tight focusing regime}. When such a structured beam is tightly focused, the longitudinal component of the electric field is enhanced, and the polarization structure forms a 3-dimensional topology. We utilized a recently introduced nanoscopic field reconstruction technique to measure all three components of the electric field as well as their relative phases. Calculating the major axes of the polarization ellipses from the measured data in 3-dimensional space reveals a ribbon-type topology with two twists {when following a circular trace} around the optical axis. We also observed the evolution of the multi-twisted polarization ribbon upon propagation through the focus.

\section{Acknowledgement}
This work was supported by the European Union's Horizon 2020 Research and Innovation Programme (Q-SORT), grant number 766970. L. M. and E. S. acknowledge financial support from the European Union Horizon 2020 program, within the European Research Council (ERC) Grant No. 694683, PHOSPhOR. F. B., R. W. B. and E. K. acknowledge the support of Canada Research Chairs (CRC) program, and Natural Sciences and Engineering Research Council of Canada (NSERC).

%Bibliography
\section*{References}
\bibliographystyle{iopart-num}
%\bibliography{ribbib.bib}

\providecommand{\newblock}{}

\end{document}